\documentclass[supplement]{ptptex}

\usepackage{graphicx}

\newcommand{\diff}[2]{\frac{\partial #1}{\partial #2}}

\title{
Description of $\eta$-distributions at RHIC energies\\ in terms of a stochastic 
model
}

\author{
Minoru \textsc{Biyajima}$^{1,2}$, Masaru \textsc{Ide}$^1$, 
Masahiro \textsc{Kaneyama}$^1$,\\ 
Takuya \textsc{Mizoguchi}$^3$ and Naomichi \textsc{Suzuki}$^4$
}

\inst{
$^1$Department of Physics, Shinshu University, Matsumoto 390-8621, Japan\\
$^2$The Niels Bohr Institute, DK-2100, Copenhagen, Denmark\\
$^3$Toba National College of Maritime Technology, Toba 517-8501, Japan\\
$^4$Matsumoto University, Matsumoto 390-1295, Japan
}

\abst{
To explain $\eta$-distributions at RHIC energies we consider the 
Ornstein-Uhlenbeck process. To account for hadrons produced in the central 
region, we assume existence of third source located there ($y \approx 0$) 
in addition to two sources located at the beam and target rapidities 
($\pm y_{\rm max} = \pm \ln\left[\sqrt{s_{\tiny NN}}/m_{\tiny
N}\right]$). This results in better $\chi^2/$n.d.f. than those for only two
sources when analysing data. 
}

\begin{document}

\maketitle

We are interested in analyses of $\eta$-distributions~\cite{Back:2001bq,Bearden:2001xw} 
at  RHIC energies by means of stochastic process, in particular the Ornstein-Uhlenbeck 
(O-U) process.\cite{Biyajima:2002at,Biyajima:2002wq} Our previous formulation 
was a simplified  approach, because we have sought an empirical formula. 
Through those studies, we have found that 1) $\eta$-distributions should be a 
sum of two or more Gaussian distributions, and 2) a  $z_r = \eta/\eta_{\rm rms}$ 
scaling holds among various centrality cuts. We have used the following 
Fokker-Planck equation
\begin{eqnarray}
  \diff{P(y,\, t)}t = \gamma \left[\diff{[yP(y,\, t)]}y + 
  \frac 12\frac{\sigma^2}{\gamma}\diff{^2P(y,\, t)}{y^2}\right]\,,
\label{eq1}
\end{eqnarray}
where $y$, $\gamma$ and $\sigma^2$ are the rapidity, the frictional coefficient 
and the variance, respectively. The solution with the initial condition 
$P(y,\, 0) = 0.5[\delta (y + y_{\rm max})+\delta (y - y_{\rm max})]$ is given as
\begin{eqnarray}
  P(y,\, y_{\rm max},\, t) &=& 
\frac 1{\sqrt{8\pi V^2(t)}}\left\{
\exp\left[-\frac{(y+y_{\rm max}e^{-\gamma t})^2}{2V^2(t)}\right]\right . 
\nonumber\\
 &&\qquad\qquad\quad\left .+ \exp\left[-\frac{(y-y_{\rm max}e^{-\gamma t})^2}{2V^2(t)}\right]\, 
 \right\}\,,
\label{eq2}
\end{eqnarray}
where $V_1^2(t) = (\sigma^2/2\gamma)p$ with $p=(1-e^{-2\gamma t})$. See 
Refs.~\citen{Wolschin:1999jy} and \citen{Morita:2002av}.

\begin{figure}
    \centerline{\includegraphics[height=7.0 cm]
                                {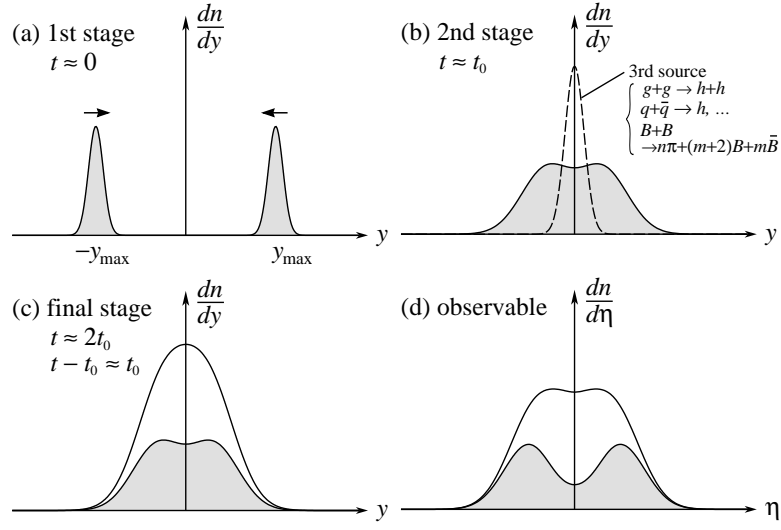}}
\caption{Description of two-step processes by O-U process.}
\label{fig1}
\end{figure}

In this
paper we
investigate a more realistic formulation, 
namely we take into account
the contribution from the central region at $y\approx 0$.\cite{Wolschin:2003zg} 
For our aim, first of all, we adopt the two-step processes by O-U process shown 
in Fig.~\ref{fig1}. 
%
%

For the third source
we assume the following $y$ distribution
\begin{eqnarray}
  P_0(y,\:t) = \frac 1{\sqrt{2\pi V_0^2(t)}}\exp\left[-\frac{y^2}{2V_0^2(t)}\right]\:,
\label{eq3}
\end{eqnarray}
where $P_0(y,\,0) = \delta(y)$ and $V_0^2(t) = (\sigma_0/2\gamma)p$. According to the physical 
picture mentioned 
above we have therefore following formula
for the normalized distribution ($\int dn/d\eta\cdot d\eta = 1$) in the $\eta$-rapidity space, 
\begin{eqnarray}
  \frac{dn}{d\eta} = J(\eta,\:m/p_t) \times r_f \times {\rm Eq.\ (\ref{eq2})} 
  + J_0\times (1-r_f)\times {\rm Eq.\ (\ref{eq3})}\,,
\label{eq4}
\end{eqnarray}
with 
$$
y = \frac 12\ln \left\{\left[\sqrt{1+(m/p_t)^2 + \sinh^2 \eta} + 
\sinh \eta\right]/\left[\sqrt{1+(m/p_t)^2 + \sinh^2 \eta} - \sinh \eta\right]\right\}\:,
$$
where the Jacobian factor $J(\eta,\:m/p_t) = \cosh \eta/\sqrt{1+(m/p_t)^2 + \sinh^2 \eta}$ 
and $J_0 = J(\eta,\:m_0/p_t)$. $r_f$ is the ratio of the weight factors among the two sources 
($c_1$) at $\pm y_{\rm max}$ and the third one ($c_0$); $r_f = 2c_1/(2c_1+c_0)$.

In concrete analyses, the masses of hadrons in the beam and target nuclei should be larger 
than those of the central region, because of the richness of baryons; 
$(m/p_t)>(m_0/p_t)$, $(m/p_t)> \delta(m/p_t)$, $(m_0/p_t) > \delta(m_0/p_t)$. The evolution 
parameter $p$ is determined at the minimum values of $\chi^2$'s, 
as in the previous 
treatments.\cite{Biyajima:2002at,Biyajima:2002wq}

\begin{figure}
    \centerline{\includegraphics[height=4.6 cm]
                                {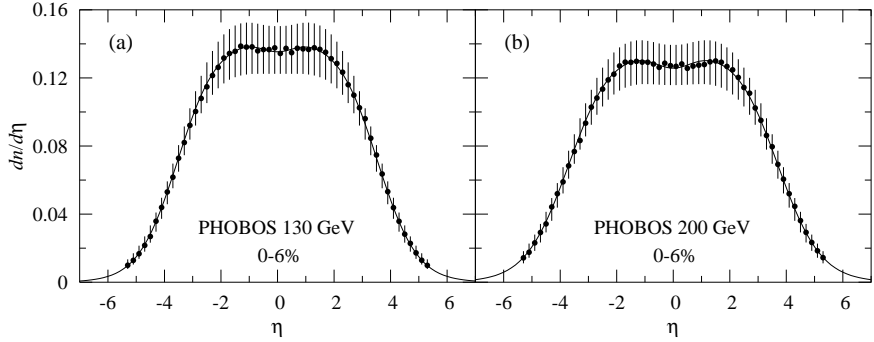}}
\caption{(a) $c_1/2:c_0:c_1/2 = 1:7:1$, $p = 0.73\pm 0.03$, $V_0^2 = 4.42\pm 0.84$, 
$V_1^2 = 1.17\pm 0.17$, 
(b) $c_1/2:c_0:c_1/2 = 1:7:1$, $p = 0.74\pm 0.33$, $V_0^2 = 5.19\pm 1.24$, 
$V_1^2 = 1.36\pm 0.23$.}
\label{fig2}
\end{figure}

Our analyses of data by PHOBOS Collaboration~\cite{Back:2001bq} at $\sqrt{s_{\tiny
NN}}=130$ GeV and $200$ GeV  are shown in Fig.~\ref{fig2} and
Table~\ref{table1}. 
As one can see there, a reasonable set of weights, which satisfy our
criteria, is $c_1/2:c_0:c_1/2=1:6:1\sim 1:10:1$. 
%
%
%
%
\begin{table}
\caption{Weights ($c_1/2:c_0:c_1/2 =1:c_0':1$) dependence of $\chi^2$'s values.}
\label{table1}
\begin{center}
\begin{tabular}{cccccccc}
\hline
& $c_0'$ & 2 & 4 & 6 & 7 & 8 & 10\\
\hline
PHOBOS & $m_0/p_t$ & 0.29$\pm$0.91 & 0.33$\pm$0.56 & 0.46$\pm$0.11 
& 0.51$\pm$0.28 & 0.55$\pm$0.25 & 0.63$\pm$0.20\\
130 GeV & $m/p_t$ & 0.55$\pm$0.18 & 0.54$\pm$0.20 & 0.68$\pm$0.15 
& 0.72$\pm$0.20 & 0.76$\pm$0.21 & 0.83$\pm$0.22\\
0-6 \% & $\chi^2/$n.d.f. & 0.79/47 & 0.71/47 & 0.64/47 & 0.62/47 & 0.62/47 & 0.65/47\\
\hline
PHOBOS & $m/p_t$ & 0.22$\pm$1.02 & 0.36$\pm$0.47 & 0.47$\pm$0.34 
& 0.51$\pm$0.29 & 0.55$\pm$0.27 & 0.61$\pm$0.23\\
200 GeV & $m/p_t$ & 0.45$\pm$0.16 & 0.54$\pm$0.24 & 0.69$\pm$0.28 
& 0.58$\pm$0.26 & 0.62$\pm$0.24 & 0.69$\pm$0.33\\
0-6 \% & $\chi^2/$n.d.f. & 0.63/47 & 0.68/47 & 0.76/47 & 0.81/47 & 0.88/47 & 1.02/47\\
\hline
\end{tabular}
\end{center}
\end{table}

\begin{figure}[h]
    \centerline{\includegraphics[height=5.6 cm]
                                {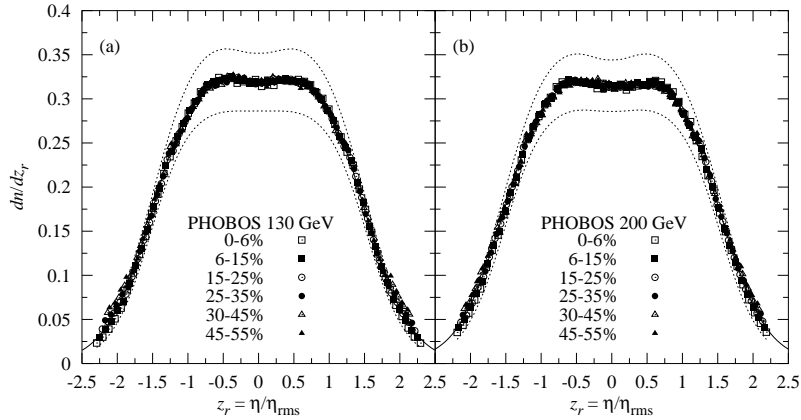}}
\caption{$z_r$ scaling by Eq.~(\ref{eq5}). (a) $\chi^2/{\rm n.d.f.} = 21.2/318$. 
(b) $\chi^2/{\rm n.d.f.} = 11.5/318$. Dotted lines mean the error bars.}
\label{fig3}
\end{figure}

Moreover, to
confirm the $z_r = \eta/\eta_{\rm rms}$ ($\eta_{\rm rms} = 
\sqrt{\langle \eta^2 \rangle}$)
scaling we
compare our theoretical formula
\begin{eqnarray}
  \eta_{\rm rms}\frac{dn}{d\eta} = \frac{dn}{dz_r} = 
  \eta_{\rm rms} \times {\rm Eq.\ (\ref{eq4})} |_{\eta =z_r\eta_{\rm rms}}
\label{eq5}
\end{eqnarray}
with data including
5 centrality cuts. Explanations of the $z_r$ scaling by means 
of Eq.~(\ref{eq5}) shown in Fig.~\ref{fig3} seem to be excellent.
%
%

\begin{figure}
    \centerline{\includegraphics[height=5.6 cm]
                                {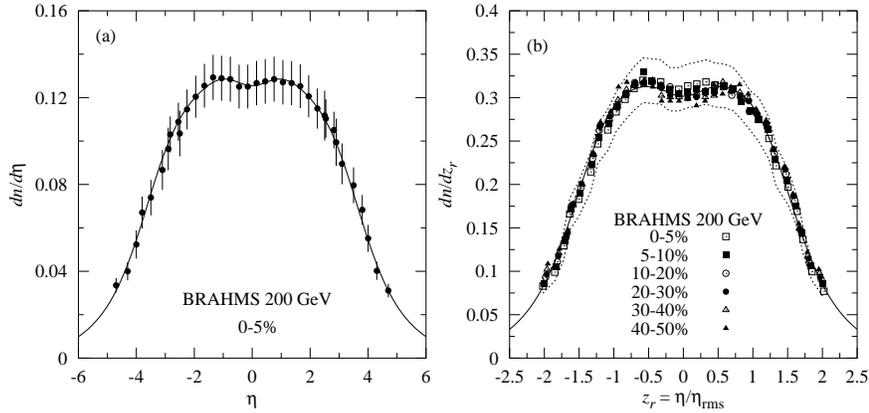}}
\caption{(a) $c_1/2:c_0:c_1/2 = 1:20:1$, $\chi^2 = 4.68/30$, $y_{max}=5.36$ (fixed), 
$p = 0.75$ (fixed), $V_0^2 = 6.42\pm 0.95$, $V_1^2 = 0.79\pm 0.23$. (b) $z_r$ scaling, 
$c_1/2:c_0:c_1/2 = 1:10:1$, $p = 0.95\pm 0.06$, $\chi^2/{\rm n.d.f.} = 29.8/186$. Dotted lines mean the error bars.}
\label{fig4}
\end{figure}

Hereafter, we analyzed
data by BRAHMS
Collaboration~\cite{Bearden:2001xw} at 200 GeV 
using
Eqs.~(\ref{eq4}) and (\ref{eq5}). In this case our criterion
mentioned above 
cannot be applied,
as seen in Table~\ref{table2}, because of fluctuation in $dn/d\eta$
and restricted data points in fragmentation regions. 
Therefore we have
assumed the evolution parameter $p=0.75$ taken from Fig.~\ref{fig2}(b)
which results in
reasonable value of $\chi^2$'s at $c_1/2:c_0:c_1/2 = 1:20:1$. The $z_r$ 
scaling shown in Fig.~\ref{fig4} holds with $c_1/2:c_0:c_1/2 = 1:10:1$ in a 
sense  of the averaged centrality cuts. 
%
%
\begin{table}
\caption{Same as Table~\ref{table1} but data by BRAHMS Collaboration at 200 GeV.}
\label{table2}
\begin{center}
\begin{tabular}{cc|ccc|ccc}
\hline
&& \multicolumn{3}{|c}{$p =$ free} & \multicolumn{3}{|c}{$p = 0.75$ (fixed)}\\
& $c_0'$ & 5 & 10 & 20 & 5 & 10 & 20\\
\hline
BRAHMS & $m_0/p_t$ & 0.57$\pm$0.33 & 0.51$\pm$0.18 & 0.56$\pm$0.16 
                   & 0.29$\pm$0.38 & 0.37$\pm$0.46 & 0.55$\pm$0.24\\
200 GeV & $m/p_t$ & 6.2$\pm$2.9 & 7.1$\pm$2.7 & 16.7$\pm$25.0 
                  & 0.00$\pm$2.15 & 0.35$\pm$1.57 & 0.78$\pm$0.66\\
0-5 \% & $p$ & 1.00$\pm$0.80 & 1.00$\pm$0.84 & 1.00$\pm$0.03 
             & 0.75 & 0.75 & 0.75\\
& $\chi^2/$n.d.f. & 4.70/30 & 4.56/30 & 3.89/30 
                  & 4.94/31 & 4.70/31 & 4.37/31\\
\hline
\end{tabular}
\end{center}
\end{table}
%
%

In conclusion, the two-step processes by O-U process is available for the 
description of $\eta$-distributions at RHIC energies.

Acknowledgements: One of authors~(M.~B.) would like to thank the 
Scandinavia-Japan Sasakawa Foundation for financial support, 2002. In addition 
this study is partially supported by a research program of Shinshu 
University, 2002. They are indebted to G.~Wilk for his reading the manuscript. 
Moreover, we would like to appreciate various conversations 
at International Workshop "Finite Density QCD" held in Nara.

%

\end{document}